\documentclass[a4paper,12pt]{iopart}
\usepackage[utf8]{inputenc}
\usepackage[T1]{fontenc}

\usepackage{amsmath,amssymb}
\usepackage{microtype}
\usepackage{array}
\usepackage{graphicx}
\usepackage[rgb]{xcolor}
\usepackage[numbers,sort&compress]{natbib}
\usepackage{hyperref}
\hypersetup{colorlinks,linkcolor=blue,citecolor=blue,urlcolor=blue}
\def\equationautorefname~#1\null{Eq.~(#1)\null}

\newcommand{\wcut}{\omega_{\mathrm{cut}}}
\newcommand{\PDS}{\mathcal{D}}

\newcommand{\bra}[1]{\langle #1 |}
\newcommand{\ket}[1]{| #1 \rangle}

\begin{document}

\title[Quantum theory of collective strong coupling of molecular vibrations with a cavity]{Quantum theory of collective strong coupling of molecular vibrations with a microcavity mode}
\author{Javier del Pino$^1$, Johannes Feist$^1$, Francisco~J.~Garcia-Vidal$^{1,2}$}
\address{$^1$Departamento de F{\'\i}sica Te{\'o}rica de la Materia Condensada and Condensed Matter Physics Center (IFIMAC), Universidad Aut\'onoma de Madrid, Madrid E-28049, Spain}
\address{$^2$Donostia International Physics Center (DIPC), E-20018 Donostia/San Sebastian, Spain}
\ead{\href{mailto:johannes.feist@uam.es}{johannes.feist@uam.es}}

\pacs{42.50.Nn, %quantum optical phenomena
71.36.+c, % Polaritons
78.66.Qn % Polymers; organic compounds
}

\begin{abstract}
We develop a quantum mechanical formalism to treat the strong coupling between an electromagnetic mode and a vibrational excitation of an ensemble of organic molecules. By employing a Bloch-Redfield-Wangsness approach, we show that the influence of dephasing-type interactions, i.e., elastic collisions with a background bath of phonons, critically depends on the nature of the bath modes. In particular, for long-range phonons corresponding to a common bath, the dynamics of the ``bright state'' (the collective superposition of molecular vibrations coupling to the cavity mode) is effectively decoupled from other system eigenstates. For the case of independent baths (or short-range phonons), incoherent energy transfer occurs between the bright state and the uncoupled dark states. However, these processes are suppressed when the Rabi splitting is larger than the frequency range of the bath modes, as achieved in a recent experiment [Shalabney \emph{et al.}, \href{http://dx.doi.org/10.1038/ncomms6981}{Nat. Commun. 6, 5981 (2015)}]. In both cases, the dynamics can thus be described through a single collective oscillator coupled to a photonic mode, making this system an ideal candidate to explore cavity optomechanics at room temperature.
\end{abstract}

% \maketitle

\section{Introduction}
The field of cavity optomechanics explores the interaction between electromagnetic radiation and the quantized mechanical motion of nano- or micro-oscillators~\cite{Gigan2006,Arcizet2006,Schliesser2006}. Recent developments promise a rich variety of applications such as precision mechanical measurements and coherent control of quantum states (see \cite{Aspelmeyer2014,Aspelmeyer2014a} for recent reviews). When the interaction becomes sufficiently large, coherent energy exchange between the optical and mechanical degrees of freedom becomes possible and the system could reach the strong coupling regime. Along this direction, a novel approach consists in coupling a cavity mode to a molecular bond vibration~\cite{Shalabney2015,Long2015}, which is in the ground state already at room temperature, without requiring any cooling. In order to achieve strong coupling, this has to be done by using an ensemble of molecules. This phenomenon has common ingredients with the collective strong coupling observed when electronic excitations of organic molecules interact with cavity~\cite{Lidzey1998,Schwartz2011,Kena-Cohen2013} or plasmonic modes~\cite{Bellessa2004}, which has been studied extensively during the last years (see \cite{Torma2015,Michetti2015} for recent reviews).

In the strong coupling with an ensemble of vibrational modes, the cavity resonance couples to a \emph{collective} superposition of the molecular vibrations, the so-called bright state, forming hybrid states called polaritons. In principle, all other superpositions of vibrational excitations (the so-called dark states) remain uncoupled. However, within this setting, a thermal bath of low-frequency rovibrational modes, which normally only introduces dephasing~\cite{Oxtoby1979,Tanimura1993}, interacts with the vibrational excitations and may influence the system dynamics. It is an open question how the dephasing mechanisms affect the strongly coupled dynamics and what role the dark states play. In particular, it is unknown whether the bright mode in such a system effectively behaves like a single isolated oscillator, which would enable the direct application of protocols developed in cavity optomechanics.

\begin{figure}[tb]
\centering
\includegraphics[width=0.95\linewidth]{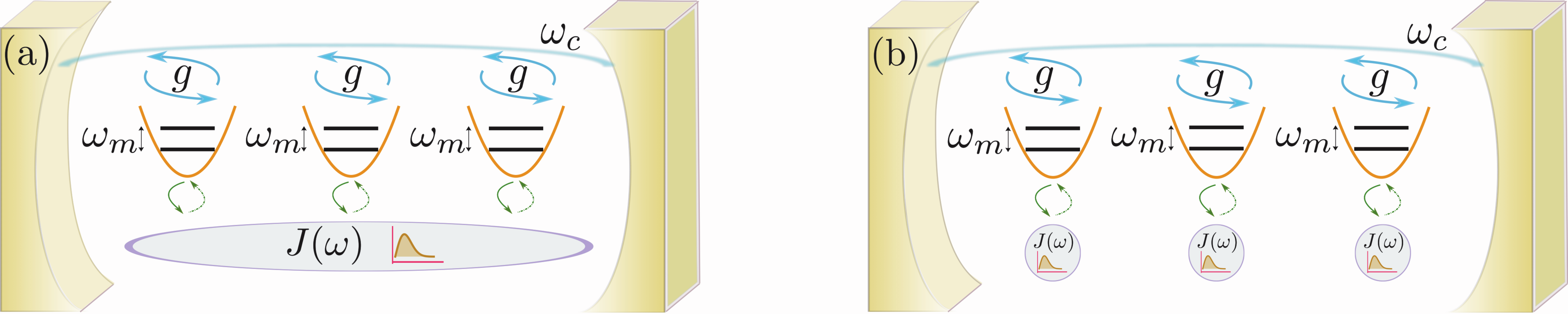}
\caption{Sketch of the model system. An ensemble of $N$ molecules interacts collectively with a cavity mode of frequency $\omega_c$. The cavity is tuned to a molecular vibrational mode. The remaining rovibrational modes are introduced by either connecting all molecules to a common bath (panel a) or by coupling each molecule with an independent bath (panel b). In both cases the harmonic bath is characterized by a spectral density $J(\omega)$.}
\label{fig:1}
\end{figure}

In this paper, we present a fully quantum theory of the phenomenon of collective strong coupling of molecular vibrations with a cavity mode. In contrast to classical transfer matrix calculations, which can be used to fit the experimental spectra~\cite{Shalabney2015,Long2015} but only provide very limited information about the incoherent dynamics in the system, our approach naturally incorporates all incoherent processes, in particular those induced by dephasing-type interactions. We describe the system using a quantum-mechanical model of a single photon mode coupled to an ensemble of harmonic oscillators representing molecular vibrations (the C=O bond stretching mode of polyvinyl acetate at an energy of $215\ $meV in the experiment~\cite{Shalabney2015}). In order to incorporate the coupling of the molecular vibrations to low-frequency rovibrational modes, we assume that the oscillators are connected to either a \emph{common} or to \emph{independent} phononic baths (see \autoref{fig:1}). We employ Bloch-Redfield-Wangsness (BRW) theory~\cite{Wangsness1953,Redfield1955} to obtain master equations describing the system dynamics under strong coupling. The final master equations only contain a few Lindblad terms, with rates determined by the product of i) the bath spectral density at the corresponding transition frequency and ii) algebraic prefactors obtained from transforming the system into its eigenstate basis. Using the specific properties of the system under study, we can evaluate all these prefactors \emph{analytically}, and are thus able to directly read off the population transfer rates between the system eigenstates. This allows detailed insight into the system dynamics and, specifically, the role of the dark states. Our results demonstrate that for large enough Rabi splitting, the bright state indeed behaves like a \emph{single} isolated oscillator and the dark states play an almost negligible role. For the case of a common bath, corresponding to long-range bath phonons, this is even true regardless of the Rabi splitting. 

\section{Theory}
\subsection{Coherent dynamics}
We model the system as a set of $N$ molecular vibrational modes coupled to a \emph{single} electromagnetic (EM) mode in a microcavity, as depicted in \autoref{fig:1}. The mirrors in the experiments~\cite{Shalabney2015,Long2015} are actually planar, and the photonic modes form a continuum, with a dispersion relation depending on the in-plane momentum $\vec{k}$. However, the assumption of a single EM mode coupling to many molecules is justified when comparing the density of EM modes with the molecule density in the experiment. The 2D density of EM modes with wavevector $k<k_{\mathrm{max}}$ (where $k=|\vec{k}|$) is $N_{\mathrm{ph}}=\frac{1}{4\pi^2}\int_0^{k_{\mathrm{max}}} 2\pi k\,\mathrm{d}k = k_{\mathrm{max}}^2/(4\pi)$. For the $j$th transversal mode in a cavity of length $L$ with background refractive index $n$, the dispersion relation is $\omega=\frac{c}{n}\sqrt{k^2 + (\frac{j\pi}{L})^2}$, so that the density of states with energy $\omega<\omega_{\mathrm{max}}$ is (here and in the following, we set $\hbar=1$)
\begin{equation}
N^{(j)}_{\mathrm{ph}}=\frac{n^2}{4\pi c^2}(\omega_{\mathrm{max}}^2 - \omega_j^2)\,,
\end{equation}
where $\omega_j=\frac{c}{n}\frac{j\pi}{L}$. We assume that modes with $j=1$ and $\omega<\omega_{\mathrm{max}} \approx \omega_1 +\Omega_R$ participate in the dynamics. Taking the parameters from the experiment of Shalabney \emph{et al.}~\cite{Shalabney2015} ($n \approx 1.41$, $\omega_1=215\ $meV, $\Omega_R=20.7\ $meV, $L\approx2\ \mu$m), this leads to an EM mode density of $N_{\mathrm{ph}}\approx4\cdot10^6\ $cm$^{-2}$. In contrast, the reported molecular density is $d\approx 8\cdot10^{21}\ $cm$^{-3}$, giving a 2D molecular density of $L d\approx 2\cdot10^{18}\ $cm$^{-2}$. There are thus on the order of $10^{12}$ molecular vibrational modes coupled to each photonic mode, and our assumption is clearly justified. 

Within the rotating wave approximation, i.e., neglecting processes that create or destroy two excitations, the coherent dynamics of the system is then governed by the Hamiltonian:
\begin{equation}
H_s=\omega_c a^{\dagger}a + \sum_{i=1}^N \omega_{i} c_{i}^{\dagger}c_{i}+\sum_{i=1}^Ng_{i}(ac_{i}^{\dagger}+a^{\dagger}c_{i})\,,
\label{eq:Hs}
\end{equation}
where $a$ is the annihilation operator for the cavity mode with frequency $\omega_c$, and $c_i$ is the annihilation operator of the optically active vibrational mode of molecule $i$, characterized by its frequency $\omega_i$ and its position $\vec{r}_i$. The cavity-oscillator interaction is given by $g_i$, which depends on the cavity electric field strength and the change of the molecular dipole moment under displacement from the equilibrium position (see \cite{Shalabney2015} for details). For simplicity, in this work we assume a regular configuration in which all the molecules are identical ($\omega_i=\omega_m$, $g_i=g$), as well as zero detuning ($\omega_c=\omega_m$). We have explicitly checked that orientational disorder ($g_i\not=g_j$) and inhomogenous broadening ($\omega_i\not=\omega_j$) do not significantly affect the results presented below.

We focus on the linear response of the system, so that we can restrict the treatment to the zero- and single-excitation subspaces\footnote{We note that since the system Hamiltonian is quadratic in the bosonic modes $\{a,c_i\}$, it can in principle be diagonalized without restriction to any excitation subspace. This leaves most of the derivation presented in the following unchanged, but introduces additional prefactors depending on the excitation numbers in the incoherent rates derived below. For simplicity, we thus allow at most one excitation.}. The $N+1$ singly excited eigenstates of $H_s$ are formed by: i) two \emph{polaritons}, $\ket{\pm}$, symmetric and antisymmetric linear combinations of the cavity mode $a^\dag\ket{0}$, with the collective \emph{bright state} of the molecular excitation, $\ket{B}=\frac{1}{\sqrt{N}}\sum_i c_i^\dagger\ket{0}$; $\ket{\pm}=\frac{1}{\sqrt{2}}(a^\dagger\ket{0}\pm\ket{B})$ and ii) the so-called \emph{dark states}, $N-1$ combinations of molecular excitations orthogonal to $\ket{B}$, which have eigenfrequencies $\omega_m$ and no mixing with the photonic mode. The eigenfrequencies of the two polariton modes are $\omega_m\pm g\sqrt{N}$, with Rabi splitting $\Omega_R=2g\sqrt{N}$. Collective strong coupling emerges when $\Omega_R$ is larger than the losses of the system. We can distinguish three types of loss mechanisms: cavity losses (rate $\kappa$), non-radiative internal losses within the molecules (rate $\gamma_{nr}$), and dephasing-type interactions. Spontaneous radiative decay is very slow (on the scale of milliseconds) due to the low transition frequencies and can be safely neglected.

\subsection{Incoherent dynamics}
We now turn to the description of the incoherent dynamics induced by the different loss mechanisms.
Whereas cavity losses and nonradiative internal molecular decay can be seen as pure decay channels and included as constant Lindblad terms when analyzing the system dynamics, dephasing-type interactions must be treated in a more detailed fashion, as we detail below. In the case analyzed in this work, these interactions are due to elastic scattering of low-frequency rovibrational bath modes with the main vibrational modes involved in strong coupling, described by the interaction Hamiltonian
\begin{equation}
H_{\phi}=\sum_{i=1}^N c_i^{\dagger}c_i \sum_{k}\lambda_{ik} (b_{ik}+b_{ik}^{\dagger})\,.
\end{equation}
The spatial extension or localization of the bath modes determines the character of the coupling. In the following, we focus on two limiting scenarios. In the first scenario, all the vibrational modes are coupled to the same \emph{common} bath (see \autoref{fig:1}a), characterized by delocalized phonons $b_{ik}=b_{k}$ with bath Hamiltonian $H^{\mathrm{com}}_b = \sum_k \omega_k b_k^\dag b_k$. In the second scenario, each molecular vibrational mode is coupled to an \emph{independent} bath characterized by on-site phonons $b_{ik}$, with bath Hamiltonian $H^{\mathrm{ind}}_b = \sum_{ik} \omega_{ik} b_{ik}^\dag b_{ik}$. In both scenarios, we assume that all molecules are identical ($\lambda_{ik}=\lambda_k$ and $\omega_{ik}=\omega_k$).

The frequently used approach of treating dephasing through a frequency-independent Lindblad superoperator acting only on the vibrational modes is not valid in our case as, within the strong coupling regime, the molecule-cavity coupling (frequency $\Omega_R$) is much faster than the correlation time of the phononic environment. Instead, the influence of the background modes has to be taken into account in the \emph{dressed} basis obtained after diagonalizing the strong-coupling interaction. The bath is completely characterized by the spectral density $J_i(\omega) = \sum_{k} \lambda_{i k}^2 \delta (\omega - \omega_{i k})$ (with $J_i(\omega)=J(\omega)$ in our case). In the following, we assume an Ohmic environment with a quadratic cutoff at frequency $\wcut$
\begin{equation}
J(\omega)=\eta\omega e^{-(\omega/\wcut)^2}, \label{eq:J}
\end{equation}
where $\eta$ is a dimensionless constant that determines the system-bath coupling strength.

If the system-bath coupling is sufficiently weak, Bloch-Redfield-Wangsness (BRW) theory~\cite{Wangsness1953,Redfield1955} can be safely applied to derive a master equation for the system dynamics~\cite{Gardiner2004,Petruccione2007}. This approach relies on the first and second \emph{Born approximation}, which calculate the effect of the system-bath coupling up to second order perturbation theory and assume that the bath state remains unmodified, i.e., the bath density matrix $\rho_b$ is time-independent (and thermally populated in the following), where $\rho_{\mathrm{total}}(t)\simeq\rho(t)\otimes\rho_b$. Additionally, since the decay of the bath correlations $\sim1/\wcut$ occurs on a much shorter time scale than the dynamics caused by the interaction with the bath, the \emph{Markov approximation} is used, disregarding all memory effects of the system-bath interaction. In the interaction picture (denoted by a tilde, $\tilde O = e^{i(H_s+H_b)t} O e^{-i(H_s+H_b)t}$), the system density operator $\tilde \rho$ then evolves according to
\begin{equation}
\partial_t\tilde{\rho} (t) = -\int\limits_{-\infty}^t \Tr_b [\tilde{H}_{\phi}(t), [\tilde{H}_{\phi}(t'), \tilde{\rho}(t)\otimes\rho_b]] \mathrm{d}t',
\label{eq:secon}
\end{equation}
where $\Tr_b$ denotes the trace over the bath degrees of freedom.

The bath-dependent part of the system-bath coupling is fully encoded in the bath \emph{correlation functions} between the modes on molecular sites $i$ and $j$, $\phi_{i,j}(\tau) = \sum_k\Tr_b[\tilde{b}_{ik}^\dag(\tau) \tilde{b}_{jk}(0) \rho_b]$. It is independent of both $i$ and $j$ for a common bath, $\phi^{\mathrm{com}}_{i,j}(\tau) =\phi(\tau)$, while for independent baths, the off-diagonal terms vanish, $\phi^{\mathrm{ind}}_{i,j}(\tau)=\delta_{i,j}\phi(\tau)$. The \emph{autocorrelation} function $\phi(\tau)$ can be expressed through the spectral density~\cite{Mahan2000},
\begin{equation}
\phi(\tau)=\int\limits_0^\infty J(\omega) \left\{[n(\omega)+1]e^{-i\omega\tau}+n(\omega)e^{i\omega\tau}\right\} \mathrm{d}\omega,
\label{eq:corre}
\end{equation}
where $n(\omega)= (e^{\beta \omega}-1)^{-1}$ is the Bose occupation factor and $\beta=1/k_BT$, with $k_B$ the Boltzmann constant and $T$ the temperature. 

We now evaluate \autoref{eq:secon} within the system eigenbasis. Before we proceed, we note that the common approach of using frequency-independent Lindblad terms to describe dephasing is equivalent to neglecting the strong coupling in the incoherent dynamics, i.e., to use only the uncoupled system Hamiltonian $\omega_ca^{\dagger}a+\omega_m\sum_i c_{i}^{\dagger}c_{i}$ in the interaction picture on the right-hand side of \autoref{eq:secon}.
When the molecule-cavity coupling is comparable to or faster than the decay of bath correlations $(\Omega_R\gtrsim\wcut)$, this is an invalid approximation,
and it is crucial to include the full system Hamiltonian when deriving the master equation to satisfy detailed balance~\cite{Carmichael1973}.

We thus proceed by expressing the system part of $H_\phi$ ($\propto c_i^{\dagger}c_i$) in terms of the dressed eigenbasis and inserting this expansion in \autoref{eq:secon}. This leads to
\begin{equation}
  \partial_t{\tilde{\rho}} = \sum_{i,j=1}^{N} \sum_{p,q,r,s} \int\limits_0^\infty \phi_{ij}(\tau) u_{ip}u_{iq}u_{jr}u_{js}
  e^{i(\omega_{pq}-\omega_{sr})t + i\omega_{sr}\tau} \left[\sigma_{rs}\tilde{\rho}(t),\sigma_{pq}\right] \mathrm{d}\tau + \mathrm{H.c.}, \label{eq:master_presec}
\end{equation}
where $\sigma_{ab}=\ket{a}\bra{b}$, $H_s \ket{a} = \omega_a \ket{a}$, $\omega_{ab}=\omega_a-\omega_b$, and the sums over $p$, $q$, $r$ and $s$ include all system eigenstates. Furthermore, $u_{ia} = \bra{a}c_i^\dagger\ket{0}$ give the overlaps between system eigenstates and vibrational mode excitations and can be chosen real. Finally, we made the substitution $t'=t-\tau$. Inserting $\phi(\tau)$ from \autoref{eq:corre} leads to integrals of the type
\begin{equation}
\int_{0}^{\infty} e^{\pm i \Delta\omega\,\tau} \mathrm{d}\tau = \pi\delta(\Delta\omega) \pm i\,\mathrm{P.V.} (\Delta\omega^{-1}),
\end{equation}
where $\mathrm{P.V.}$ denotes the Cauchy principal value---we neglect these imaginary parts as they only induce small energy shifts (\emph{Lamb shifts}) that can be reabsorbed in the coherent dynamics, and arrive to
\begin{equation}
  \partial_t\tilde\rho = \sum_{i,j=1}^{N} \sum_{p,q,r,s} S_{ij}(\omega_{sr}) u_{ip}u_{iq}u_{jr}u_{js}
  \left\{e^{i(\omega_{pq}-\omega_{sr})t} \left[\sigma_{rs}\tilde{\rho}(t),\sigma_{pq}\right] + \mathrm{H.c.}\right\},
  \label{eq:master_presec2}
\end{equation}
with the bath noise-power spectrum
\begin{equation}
S(\omega) = \begin{cases}\pi J(\omega) [n(\omega)+1] & \omega\geq 0\\
  \pi J(-\omega) n(-\omega) & \omega<0\end{cases}
\end{equation}
For a common bath, $S^{\mathrm{com}}_{ij}(\omega)=S(\omega)$, while for independent baths, $S^{\mathrm{ind}}_{ij}(\omega)=\delta_{i,j} S(\omega)$. Note that $S(0) \equiv \lim\limits_{\omega\to0^+} S(\omega) = \lim\limits_{\omega\to0^-} S(\omega)$ is well-defined if $\lim\limits_{\omega\to0}\frac{J(\omega)}{\omega}$ exists.

In the resulting master equation, the terms in which $\omega_{pq}$ differs from $\omega_{sr}$ oscillate as a function of time $t$, which can lead to a violation of the positivity of $\rho$ for long times. If the timescale $\tau_\phi$ of the bath-induced system dynamics is much slower than the coherent dynamics, i.e., $\tau_\phi \gg \tau_{\mathrm{SC}}$, these terms can be removed by averaging the master equation over a time short compared to $\tau_\phi$, but long compared to $\tau_{\mathrm{SC}}$~\cite{Dumcke1979,Spohn1980}. 
We can estimate $\tau_\phi \sim 1/\gamma_\phi$, with $\gamma_\phi$ the bare-molecule dephasing rate, for which the total bare-molecule width presents an upper bound, $\gamma_\phi \leq \gamma=3.2\ $meV~\cite{Shalabney2015}. 
As a consequence, only the \emph{secular} terms $\omega_{pq}=\omega_{sr}$ persist. This \emph{secular approximation}\footnote{The secular approximation is also sometimes called the \emph{rotating wave approximation}, but should not be confused with the rotating wave approximation performed in the system Hamiltonian in \autoref{eq:Hs}.} results in a master equation where populations and coherences are decoupled. We note that the secular approximation aids in the interpretation of the different terms that are obtained, but is not required and indeed only used in some of the results shown in the following.

\begin{table}
  \centering
  \begin{tabular}{| >{\centering\hspace{-5pt}}p{0.1\linewidth} | >{\centering}p{0.3\linewidth} | >{\centering\arraybackslash}p{0.15\linewidth}|}
    \hline
    $\phantom{-}\omega_{sr}$ & $\{pq,rs\}$ & Label for $\gamma_{rs}$\\
    \hline\hline
    $\phantom{-}\Omega_{R}$ & $\{+-,-+\}$ & $\gamma_e$ \\
    \hline
    $-\Omega_{R}$ & $\{-+,+-\}$ & $\gamma_a$ \\
    \hline
    $\displaystyle \phantom{-}\frac{\Omega_{R}}{2}$ & $\begin{array}{c}
                                                         \{+d,d'+\}, \{d-,-d'\}\\
                                                         \{+d,-d'\}, \{d-,d'+\}
                                                       \end{array}$ & $\Gamma_e$ \\
    \hline
    $\displaystyle-\frac{\Omega_{R}}{2}$ & $\begin{array}{c}
                                              \{-d,d'-\}, \{d+,+d'\}\\
                                              \{-d,+d'\}, \{d+,d'-\}
                                            \end{array}$ & $\Gamma_a$ \\
    \hline
    $\phantom{-}0$ & $\begin{array}{c}
                        \{++,++\}, \{--,--\}\\
                        \{dd',d''d'''\}
                      \end{array}$ & $\gamma_\phi$ \\
    \hline
  \end{tabular}
  \protect\caption{Secular terms in the master equation and the labels for the associated ``bare'' rates $\gamma_{rs} = 2S(\omega_{sr})$. The dark states are labeled with $d\in 1,2,...,N-1$.}\label{table:deph}
\end{table}

The secular terms are enumerated in \autoref{table:deph}, which lists the states $\{pq,rs\}$ connected by transitions with frequency $\omega_{sr}$, as well as the labels we assign to the ``bare'' rates $\gamma_{rs} = 2S(\omega_{sr})$, which do not contain the algebraic prefactors. The definition is chosen so that for terms where $pq=sr$, we obtain
\begin{equation}
  S(\omega_{sr})  \{[\sigma_{rs}\tilde{\rho}(t),\sigma_{sr}] + \mathrm{H.c.}\} = \gamma_{rs} \mathcal{L}_{\sigma_{rs}}[\rho] \,,
\end{equation}
where $\mathcal{L}_X[\rho]=X\rho X^{\dagger} - \frac12\left\{X^{\dagger}X,\rho\right\}$ is a standard Lindblad superoperator. Positive frequencies $\omega_{sr}>0$ correspond to phonon emission where the system transitions from a higher- to a lower-energy state, while negative frequencies $\omega_{sr}<0$ correspond to phonon absorption. We obtain secular terms connecting the two polaritons ($\gamma_e=2S(\Omega_R)$, $\gamma_a=2S(-\Omega_R)$\footnote{The same $\gamma_a,\gamma_e$ are obtained under strong classical driving of a single two-level system~\cite{Eastham2013}.}), terms connecting the polaritons with the dark modes ($\Gamma_e=2S(\Omega_R/2)$, $\Gamma_a=2S(-\Omega_R/2)$), and terms connecting states with the same energy ($\gamma_\phi=2S(0)$, equal to the bare-molecule dephasing rate). The latter give pure dephasing for the polaritons $\ket{+}$, $\ket{-}$, but produce coupling between populations and coherences for all dark states.

The final master equations are obtained by using the properties of the basis transformation matrix $u_{ia}$ to evaluate the algebraic prefactors $\sum_i u_{ip}u_{iq}u_{ir}u_{is}$ (independent baths) and $\sum_{i,j} u_{ip}u_{iq}u_{jr}u_{js}$ (common bath) in \autoref{eq:master_presec2}. Specifically, we use that i) the polariton-vibrational mode overlaps are given by $u_{i\pm}=\pm\frac{1}{\sqrt{2N}}$, ii) that dark states are orthogonal to each other $(\sum_i u_{id}u_{id'} = \delta_{d,d'})$, and iii) that dark states are orthogonal to the polaritons $(\sum_i u_{id} = 0)$. Note that $i$ and $j$ in the sums are molecule indices (not including the cavity mode), so that $\sum_i u_{ip}u_{iq}=\delta_{p,q}$ is generally not true.

\begin{figure}[tb]
\centering
\includegraphics[width=0.95\linewidth]{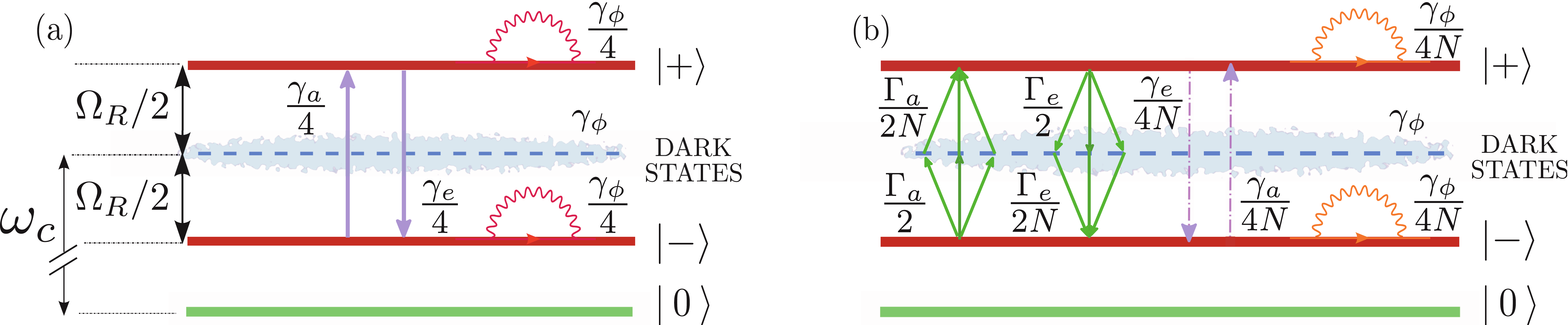}
\caption{Illustration of the different decay and dephasing mechanisms that emerge from our modeling for the two dephasing scenarios:
common bath (panel a) or independent baths (panel b). The arrows indicate decay processes while wavy lines represent elastic events that produce dephasing.
The fuzzy halo around the dark states indicates excitation transfer and dephasing acting within this manifold.}
\label{fig:2}
\end{figure}

This procedure gives the final secularized master equation for the density operator, given in the Schrödinger picture below. For a common bath with $\phi_{ij}(\tau)=\phi(\tau)$, we find that many terms vanish because of the orthogonality relations, giving
\begin{align}
\partial_t\rho &= - i[H_s,\rho] + \frac{\gamma_{a}}{4}\mathcal{L}_{\sigma_{+-}}[\rho] +\frac{\gamma_{e}}{4}\mathcal{L}_{\sigma_{-+}}[\rho]
                 + \frac{\gamma_{\phi}}{4}\sum_{p=+,-}\mathcal{L}_{\sigma_{pp}}[\rho] + \gamma_{\phi} \mathcal{L}_{\PDS}[\rho] \,,\label{eq:master_common}
\end{align}
where $\PDS=\sum_d \sigma_{dd}$ is the projector into the dark-state subspace.
The Lindblad terms $\mathcal{L}_X[\rho]$ correspond to incoherent excitation transfer between system eigenstates and are depicted schematically in \autoref{fig:2}a. A phonon of frequency $\Omega_R$ may be emitted transferring an excitation from the upper polariton $\ket{+}$ to the lower polariton $\ket{-}$, with rate $\gamma_e/4$. Phonon absorption occurs analogously, with characteristic rate $\gamma_a/4$. Furthermore, the polaritons undergo pure dephasing with rate $\gamma_\phi/4$. The last term corresponds to bare-molecule dephasing for a common bath, but projected into the degenerate dark-state subspace (using $\PDS \sum_i c_i^\dag c_i \PDS = \PDS$). Remarkably, for the case of a common bath, i.e., long-range bath phonons, the dark states are completely decoupled from the polaritons and the bright state behaves identically to a \emph{single} oscillator interacting with the cavity field.

Turning to the case of independent baths, $\phi_{ij}(\tau)=\delta_{i,j}\phi(\tau)$, we instead find
\begin{subequations}\label{eq:master_independent}
\begin{align}
\partial_t\rho&=- i[H_s,\rho] + \frac{\gamma_{a}}{4N}\mathcal{L}_{\sigma_{+-}}[\rho]+\frac{\gamma_{e}}{4N}\mathcal{L}_{\sigma_{-+}}[\rho] \label{eq:master:a}\\
              +& \frac{\Gamma_a}{2N}\sum_d (\mathcal{L}_{\sigma_{d-}}[\rho] + \mathcal{L}_{\sigma_{+d}}[\rho]) 
                 + \frac{\Gamma_e}{2N}\sum_d (\mathcal{L}_{\sigma_{d+}}[\rho] + \mathcal{L}_{\sigma_{-d}}[\rho]) \label{eq:master:b}\\
              +& \frac{\Gamma_a}{4N}\sum_d \left(\left[\sigma_{d-}\rho,\sigma_{d+}\right]-\left[\sigma_{+d}\rho,\sigma_{-d}\right]+\mathrm{H.c.}\right) \label{eq:master:c}\\
              +& \frac{\Gamma_e}{4N}\sum_d \left(\left[\sigma_{d+}\rho,\sigma_{d-}\right]-\left[\sigma_{-d}\rho,\sigma_{+d}\right]+\mathrm{H.c.}\right) \label{eq:master:d}\\
              +& \frac{\gamma_{\phi}}{4N}\sum_{p=+,-}\mathcal{L}_{\sigma_{pp}}[\rho]
                 + \gamma_{\phi} \sum_{i} \mathcal{L}_{\PDS c_i^\dagger c_i \PDS}[\rho]\,, \label{eq:master:e}
\end{align}
\end{subequations}
which now includes excitation transfer between the polaritons and dark states ($\Gamma_a$, $\Gamma_e$), driven by phonons of frequency $\Omega_R/2$. While the individual terms are strongly suppressed by the prefactor $1/N$ (with $N\sim10^{12}$ in the experiments), excitation transfer \emph{into} the dark states still occurs efficiently due to the sum over $N-1$ dark states $d$. For $N\to\infty$, the \emph{total} population transfer from the polaritons to the dark states thus occurs with rates $\Gamma_a/2$ and $\Gamma_e/2$, as shown in \autoref{fig:2}b.
On the other hand, pure dephasing of the polaritons and direct transitions between them through absorption or emission of phonons of frequency $\Omega_R$ play a negligible role within this dephasing scenario, as their rates scale as $1/N$.
Additional terms couple between different polariton-dark state coherences (Eqs.~(\ref{eq:master:c}), (\ref{eq:master:d})), without affecting the populations. Finally, the second term in \autoref{eq:master:e} again corresponds to a bare-molecule Lindblad dephasing term that has been restricted to act only within the degenerate dark-state subspace. For $N\to\infty$, \autoref{eq:master_independent} simplifies to
\begin{align}\label{eq:master_independent2}
\partial_t\rho&=- i[H_s,\rho] + \frac{\Gamma_a}{2} \bar{\mathcal{L}}_{\PDS-}[\rho] + \frac{\Gamma_e}{2} \bar{\mathcal{L}}_{\PDS+}[\rho] 
                + \gamma_{\phi} \sum_{i} \mathcal{L}_{\PDS c_i^\dagger c_i \PDS}[\rho]\,,
\end{align}
where $\bar{\mathcal{L}}_{\PDS\pm}= \frac{1}{N-1} \sum_d \mathcal{L}_{\sigma_{d\pm}}$ is an averaged Lindblad superoperator inducing equal population transfer from a polariton to all dark states. For large $N$, the dark states thus act like a sink, and any population transferred to them is trapped and does not further participate in the polariton dynamics.

As is clear from the expressions for the different decay rates, $\Omega_R$ is the parameter that controls the interaction between the \emph{dressed} excitation and the phonon bath. Hence the importance of these decoherence mechanisms is dictated by the spectral density evaluated at $\Omega_R$ or $\Omega_R/2$. This is in clear contrast to what we would obtain by treating the interaction $H_{\phi}$ in the \emph{uncoupled} basis. In that case, the Markov approximation would have resulted in ``standard'' Lindblad terms $\gamma_{\phi}\mathcal{L}_{\sum_i c_i^\dagger c_i}(\rho)$ (common bath) or $\gamma_{\phi}\sum_i\mathcal{L}_{c_i^\dagger c_i}(\rho)$ (independent baths). In both cases, the dephasing interaction would be totally controlled by just the zero-frequency limit of the spectral density, resulting in an overestimation of the rates $\gamma_{a,e}$ and $\Gamma_{a,e}$. This fact emphasizes the key importance of deriving Lindblad terms in the strongly coupled basis when considering non-flat reservoirs~\cite{Carmichael1973}.

Apart from the decay and dephasing mechanisms induced by the dephasing-type interactions, which conserve the number of excitations, the excited states may decay through nonradiative molecular decay and cavity losses into the ground state $\ket{0}$. These decay process, which we have neglected in the theory up to now, could be described within the same framework by dissipative coupling with thermal baths (e.g., the photonic modes outside the cavity). However, as the energy shifts induced by the strong coupling are small compared to the transition frequency ($\Omega_R\ll\omega_m$), we can include them through Lindblad terms for the bare cavity and molecular vibrational modes. After removing nonsecular terms in the eigenstate basis, this gives new Lindblad terms with decay rates of $(\gamma_{nr}/2+\kappa/2)$ for the two polaritons and $\gamma_{nr}$ for the dark states.

\section{Results}
In the following, we apply our theoretical framework to the experimental results of Shalabney \emph{et al.}~\cite{Shalabney2015}. For the vibrational mode of the \emph{bare} molecules, they report a linewidth of $\gamma=3.2\ $meV, with negligible inhomogeneous broadening. This linewidth has contributions from nonradiative decay and dephasing, $\gamma=\gamma_{nr}+\gamma_{\phi}$, which can not be distinguished in the absorption spectrum. Although direct information about the weights of non-radiative and dephasing channels is thus not available, dephasing is in general expected to provide a significant contribution for vibrational transitions~\cite{Oxtoby1979,Tanimura1993}. Therefore, in our calculations we will use a factor $f$ [$\gamma_\phi=f\gamma$, $\gamma_{nr}=(1-f)\gamma$] to measure the relative importance of the two channels. In this way, the factor $\eta$ in \autoref{eq:J}, which quantifies the strength of the system-bath coupling, is simply given by $\eta=f \gamma/(2\pi k_B T_0)$ where $T_0=300\ $K. The cut-off frequency for the thermal bath of low-frequency rovibrational excitations is chosen as $\wcut=6\ $meV, corresponding to the range of low-frequency phonon modes in the system~\cite{Shalabney2015}. Finally, we use a cavity loss rate of $\kappa=17\ $meV as estimated in~\cite{Shalabney2015} through fitting of the transmission spectrum in the strong coupling regime.

\begin{figure}[tb]
\centering
\includegraphics[width=0.95\linewidth]{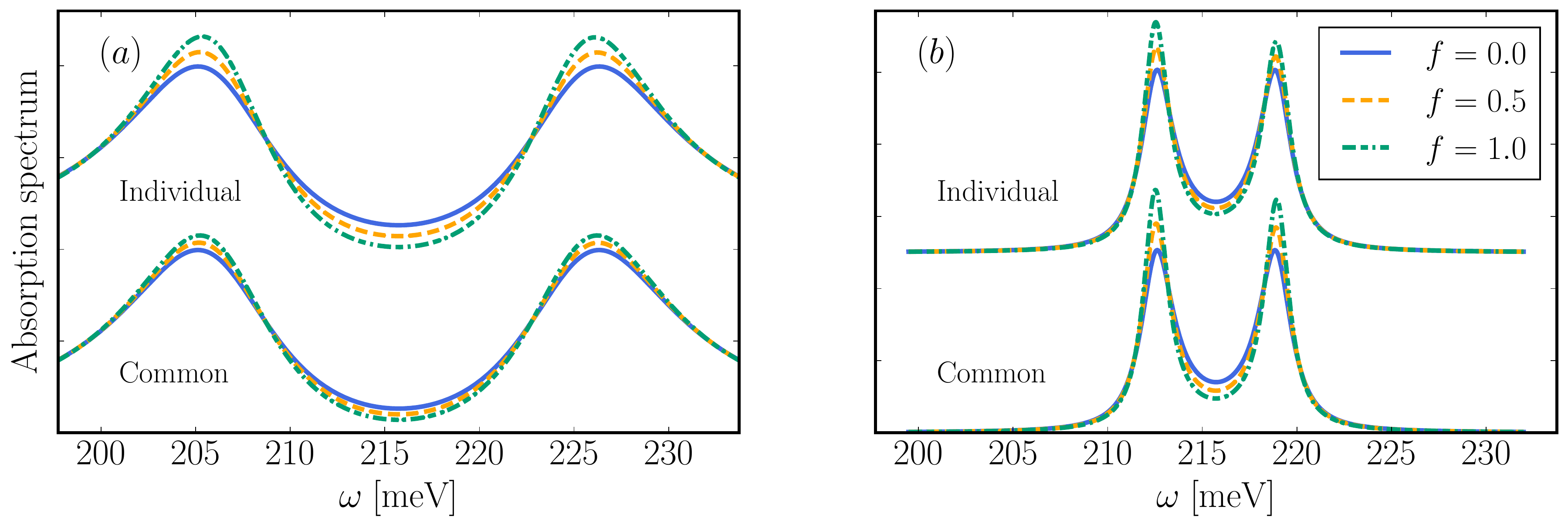}
\caption{Absorption spectra for the two possible dephasing scenarios as a function of $f$. In both panels the two sets of curves are offset for a better visualization. In panel (a), the parameters used for the calculations are those reported in Ref.~\cite{Shalabney2015} (see text), whereas in panel (b), the Rabi splitting and the cavity decay rate are reduced to $\Omega_R=6.5\ $meV and $\kappa=1\ $meV, respectively.}
\label{fig:3}
\end{figure}

We calculate the absorption spectra by introducing a weak driving term, $H_d(t)\sim ae^{-i\omega t}+\mathrm{H.c.}$, which coherently pumps the cavity mode. The density matrix in the steady state, $\rho_{ss}$, can be calculated in the frame rotating with the driving frequency $\omega$. In this frame, $H_d$ is time-independent, but the system frequencies are shifted and the density matrix $\rho_{ss}$ depends on $\omega$. The absorption spectrum is then obtained as $A(\omega)\propto\Tr[\rho_{ss}(\omega)a]$. For the numerical implementation, we employ the open-source QuTiP package~\cite{Johansson2013}.

\autoref{fig:3}a depicts the theoretical absorption spectra for the parameters reproducing the experimental situation, with Rabi splitting $\Omega_R=20.7\ $meV, for the two possible dephasing scenarios analyzed in this work (common or independent baths). Different values of $f$ are tested: $f=0$, $f=0.5$ and $f=1$. Both approaches coincide in the limit $f=0$ (no dephasing), but their behaviour differs for non-zero $f$, as can be inferred from the different decay rates as rendered in \autoref{fig:2}. The Rabi splitting is much larger than the range of low-frequency vibrations ($\Omega_R\gg\wcut$) and, hence, all terms connecting levels with different energies ($\gamma_a,\gamma_e,\Gamma_a,\Gamma_e$) are essentially zero. The widths of the polaritons in the absorption spectrum (for $N\to\infty$) are simply given by $\gamma_\pm = \frac{\kappa+\gamma_{nr}}{2} + \frac{\gamma_\phi}{4} = \frac{\kappa}{2}+(1-\frac{f}{2})\frac{\gamma}{2}$ for a common bath, and $\gamma_\pm = \frac{\kappa+\gamma_{nr}}{2} = \frac{\kappa}{2}+(1-f)\frac{\gamma}{2}$ for independent baths.

This also implies that for both bath scenarios and within the experimental conditions reported in~\cite{Shalabney2015,Long2015}, the ensemble of vibrational modes behaves effectively as just a single collective molecular oscillator, the bright state, coupled to the cavity mode. 
Consequently, the dark modes are effectively decoupled from the system dynamics under external driving.
We note here that the fit used to extract the cavity width $\kappa$ from the experimental data in~\cite{Shalabney2015} is performed under strong coupling (using a transfer matrix method). Thus, the change of linewidths predicted by our model would already be present in the observed spectra, and is consequently absorbed in the extracted cavity linewidth. This unfortunately precludes a direct test of our model based on the experimental absorption data, for which the \emph{bare} cavity linewidth would need to be available.

Furthermore, it is interesting to note that in the case of individual baths, the dephasing contributions to the polariton modes are completely suppressed, analogous to the well-known suppression of inhomogeneous broadening under strong coupling~\cite{Houdre1996}. This can be understood by the fact that dephasing and inhomogeneous broadening are closely related, corresponding respectively to temporal fluctuations or to a static distribution of the transition energies. 

\begin{figure}[tb]
\centering
\includegraphics[width=0.95\linewidth]{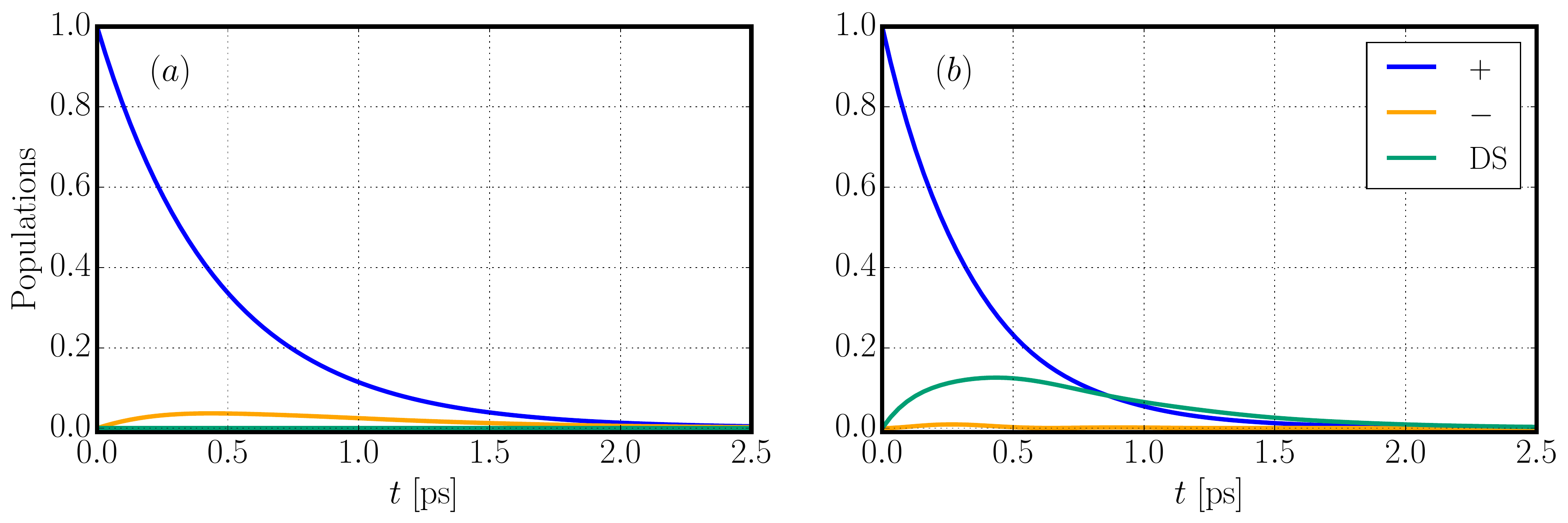}
\caption{Population dynamics starting from the initially excited upper polariton $\ket{+}$ for the two possible dephasing scenarios: (a) Common bath and (b) independent baths. The parameters are the same as in \autoref{fig:3}(b), with $f=0.5$ and $N=100$. Each panel shows the upper polariton ($+$) and lower polariton ($-$) populations, as well as the total population in all dark states (DS), given by $\Tr(\PDS\rho)$.}
\label{fig:4}
\end{figure}

The dark modes could play a bigger role in a situation with smaller Rabi splitting, for which, in order to still achieve strong coupling, the cavity losses also need to be reduced as compared to the experiments (e.g., by using thicker mirrors). The resulting absorption spectra are shown in \autoref{fig:3}b, for the same parameters as in \autoref{fig:3}a but now with $\kappa=1\ $meV and $\Omega_R=6.5\ $meV. In both bath scenarios, a slight asymmetry between upper and lower polariton is now noticeable, as the emission of phonons from the upper polariton $(\propto n(\omega)+1)$ is more likely than the absorption of phonons in the lower polariton $(\propto n(\omega))$. However, as the involved transition frequencies are smaller than the thermal energy $(k_BT=25.9\ $meV for $T=300\ $K), the thermal occupation $n(\omega)$ is significant and the rates for phonon emission and absorption are comparable. Furthermore, although the phenomenology in the observed absorption spectra is quite similar in both dephasing scenarios, with a reduction of the polariton linewidths for increasing dephasing $f$, the underlying physics are now quite distinct. This is demonstrated clearly when inspecting the population dynamics, shown in \autoref{fig:4} for the case of the upper polariton being initially excited. For the common bath, shown in \autoref{fig:4}(a), the dark states are completely decoupled from the dynamics and population is only transferred between the polaritons. In contrast, in the case of independent baths, shown in \autoref{fig:4}(b), the population primarily is transferred from the upper polariton to the dark states, which then decay through nonradiative losses. 

\section{Conclusions}
In summary, by using a fully quantum framework we have studied in detail the phenomenon of collective strong coupling when an ensemble of molecular vibrational modes interacts with a cavity electromagnetic mode, as realized experimentally in two recent papers~\cite{Shalabney2015,Long2015}. We have demonstrated that dephasing-type interactions with a thermal bath of background modes in such a system have to be treated beyond the usual Lindblad approximation in order to represent the effects of the spectral density of bath modes correctly. We have investigated two ``extreme'' scenarios for the bath, with either a common bath for all molecules, or independent baths for each molecule. For the experimentally relevant parameters, we find that the dark modes are almost totally decoupled from the polaritons in both scenarios, and the bright state behaves almost like a single isolated oscillator. Our findings thus suggest that this type of system is an ideal and simple platform to explore the exciting possibilities of cavity optomechanics at room temperature.

\section*{Acknowledgments}
This work has been funded by the European Research Council (ERC-2011-AdG proposal No. 290981), by the European Union Seventh Framework Programme under grant agreement FP7-PEOPLE-2013-CIG-618229, and the Spanish MINECO under contract MAT2011-28581-C02-01.

% needed for natbib
\newcommand\newblock{\hskip .11em plus .33em minus .07em} % important line
% make text smaller in bibliography
\def\bibfont{\footnotesize}

\bibliographystyle{apsrev4-1}
\nocite{apsrev41Control}
\bibliography{extrabibopts,references}

\end{document}